%% file: moriond.tex
\documentclass{moriond}

\def\Journal#1#2#3#4{{#1} {\bf #2}, #3 (#4)}

\def\PRD{{\em Phys. Rev.} D}

\def\APJ{{\em ApJ}}
\def\PRR{\em Phys. Rev. Research}

\def\be{\begin{equation}}
\def\ee{\end{equation}}
\def\bea{\begin{eqnarray}}
\def\eea{\end{eqnarray}}

\input{macros}

\usepackage{enumerate}
\usepackage{tikz}
\usepackage{enumitem}
\usepackage{booktabs}
\usepackage{xspace}
\usepackage{subcaption} % for subfigure
\graphicspath{figs/}
\usepackage[justification=justified,singlelinecheck=false,font=small]{caption}
\usetikzlibrary{arrows,positioning}

\newcommand{\Photo}{}

\begin{document}
\vspace*{4cm}
\title{ CMB bounds on primordial black holes with dark matter mini-halos:\\ the role of radiative feedback}

\author{ F. Scarcella }

\address{Laboratoire Univers et Particules de Montpellier,\\
CNRS \& Université de Montpellier  ,\\
34095 Montpellier, France}

\maketitle\abstracts{
Observations of the cosmic microwave background constrain the abundance of primordial black holes, as these would accrete gas and inject energy into the cosmological medium. We have revisited these constraints, taking into account the local heating and ionisation of the gas around the black holes. While constraints for \textit{naked} black holes are not significantly affected, bounds including dark matter mini-halos are drastically relaxed.
This result suggests that previous analysis may have significantly overestimated the role of dark matter mini-halos in boosting the accretion rates.}

\section{Constraints on the primordial black hole abundance}\label{sec:intro}

During the cosmic dark ages, the period between recombination and reionisation, the matter component of the Universe is mostly in the form of neutral hydrogen gas. If primordial black holes (PBHs) are present at this time, they are expected to accrete the gas, emitting high-energy radiation. This energy is deposited into the medium, affecting the  propagation of photons of the cosmic microwave background~(CMB) and thus allowing to constrain the abundance of PBHs.
Furthermore, if PBHs make up only a fraction of the dark matter~(DM), the rest of the DM is expected to cluster around them
in structures with steep density profiles, which we refer to as DM~\emph{mini-halos}. It has been claimed that the contribution of these mini-halos to the gravitational potential around the PBH can boost accretion rates -- hence the rates of energy injection -- by orders of magnitude.

In a recent work~\cite{us}, we revisited CMB constraints taking into account the effect of radiative feedback in the accretion process, \emph{i.e.} the local heating and ionisation of the gas in the vicinity of the BH. Our analysis is based on the model of Park and Ricotti~\cite{PR} (PR model), an analytical model which is which is backed up by hydro-dynamical simulations.

\begin{figure}[t]
\centering
\includegraphics[width=0.7\linewidth]{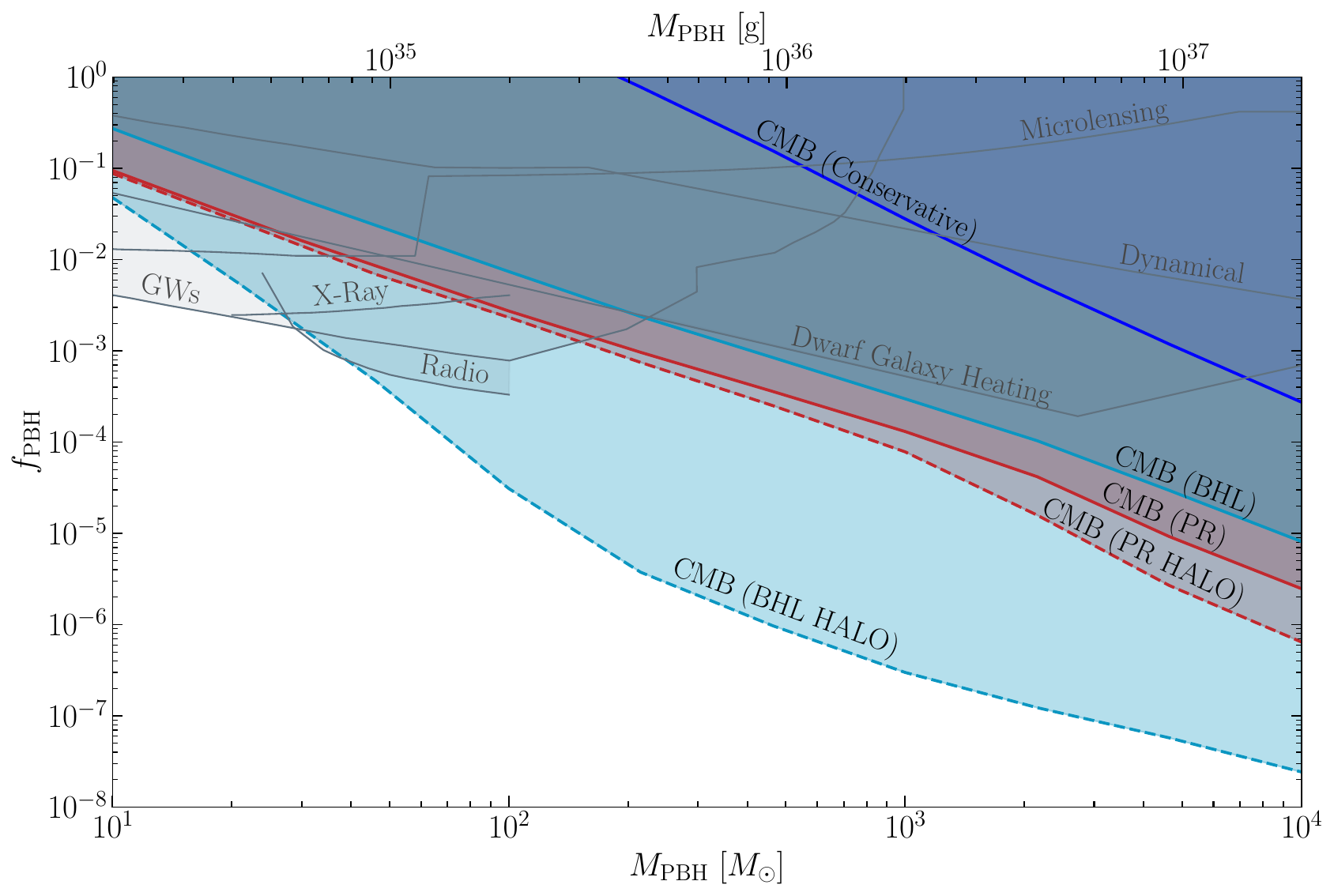}
\caption{Constraints on $\fPBH$, the fraction of DM in the form of PBHs, as a function of the PBH mass, assuming a monochromatic mass function. Light blue: BHL accretion (\textit{``previous works''}) {with either} DM halos included (dashed) {or not included} (solid). Red: PR accretion (\textit{``our work''}) {with either} DM halos included (dashed) or not included (solid). See Agius et al.~\protect\cite{us} for details.}
\label{fig:constraints}
\end{figure}

Our main results are shown in Fig.~\ref{fig:constraints}. %details of the analysis can be found in Agius et al.~\cite{us}. 
In red, bounds previously proposed in the literature; our constraints appear in blue. 
The dashed (continuous) lines correspond to accretion with (without) the contribution of DM mini-halos.
Without the contribution of the DM mini-halos, our results do not differ significantly from previous ones. The reason for this non-trivial result, and its limitations, are discussed in Agius et al.~\cite{us}

On the other hand, when DM mini-halos are included in the picture, the difference is striking. While previous works found a tightening of the bound by up to three orders of magnitude,~\cite{Serpico}
we found that radiative feedback prevents the DM mini-halos from significantly boosting the accretion rates and affecting the constraint. While we have obtained this result in the context of a particular accretion model, I argue in the following that it has more general implications, as it is driven only by the local heating of gas around the accreting PBH.

\section{Modelling the rate of gas accretion onto a black hole}\label{sec:accretion}

According to the textbook Bondi--Hoyle--Littleton (BHL) accretion model,~\cite{Bondi,HL} the accretion rate onto an isolated black hole (BH) of mass $M$ moving at speed $\vrel$ relative to its ambient medium, the latter characterised by density $\rho$ and sound speed $c_{\mathrm s}$, can be expressed as
\begin{equation}
\label{eq:Mdot}
\Mdot = \, \lambda \, 4 \pi  \rho \, \veff \, \rB^2  \, ,
\end{equation}
where $\lambda $ is a constant suppression factor~\footnote{included to match 
observations, typical values are $\lambda \sim 10^{-2}-10^{-3}$ } and the effective velocity is given by $\veff = \left( \cs^2 + \vrel^2 \right)^{1/2}$; $\rB$ is the Bondi radius, \emph{i.e.} the radius $r$ which satisfies
\begin{equation} 
\label{eq:bondiradius}
\veff ^2 = \frac{G M}{r} \,.
\end{equation} 
The Bondi radius represents an effective cross section for the accretion process. It can be regarded as the radius at which the escape velocity from the BH matches the characteristic velocity of the gas, $\veff$. This effective velocity combines the bulk velocity of the gas $\vrel$ and its sound speed $\cs$, corresponding to two sources of kinetic energy that can allow the gas particles to escape the BH potential: the bulk flow of the gas and its temperature. An increase in temperature hence leads to a reduction of the  Bondi radius and a suppression of the accretion rate. Physically, this corresponds to pressure  preventing the infall of the gas onto the BH.

Park and Ricotti~\cite{PR} performed hydro-dynamical simulations of gas accretion onto a moving BH. These showed the formation of a hot, ionised region around the BH, separated from the neutral medium by an ionisation front. They found that the observed accretion rates could be matched by applying the BHL formula (Eq.~\ref{eq:Mdot}) within the ionised region. The density, bulk velocity and temperature of the ionised gas differ from those of the neutral medium. Importantly for this discussion, its temperature is significantly higher, leading to a suppression of the size of the Bondi radius (which remains contained within the ionised region). See~\cite{us,PR} for details.

\section{The role of dark matter halos and the impact of local heating}\label{sec:DMhalos}

\begin{figure}[t!]
\centering
\includegraphics[width=.9\columnwidth]{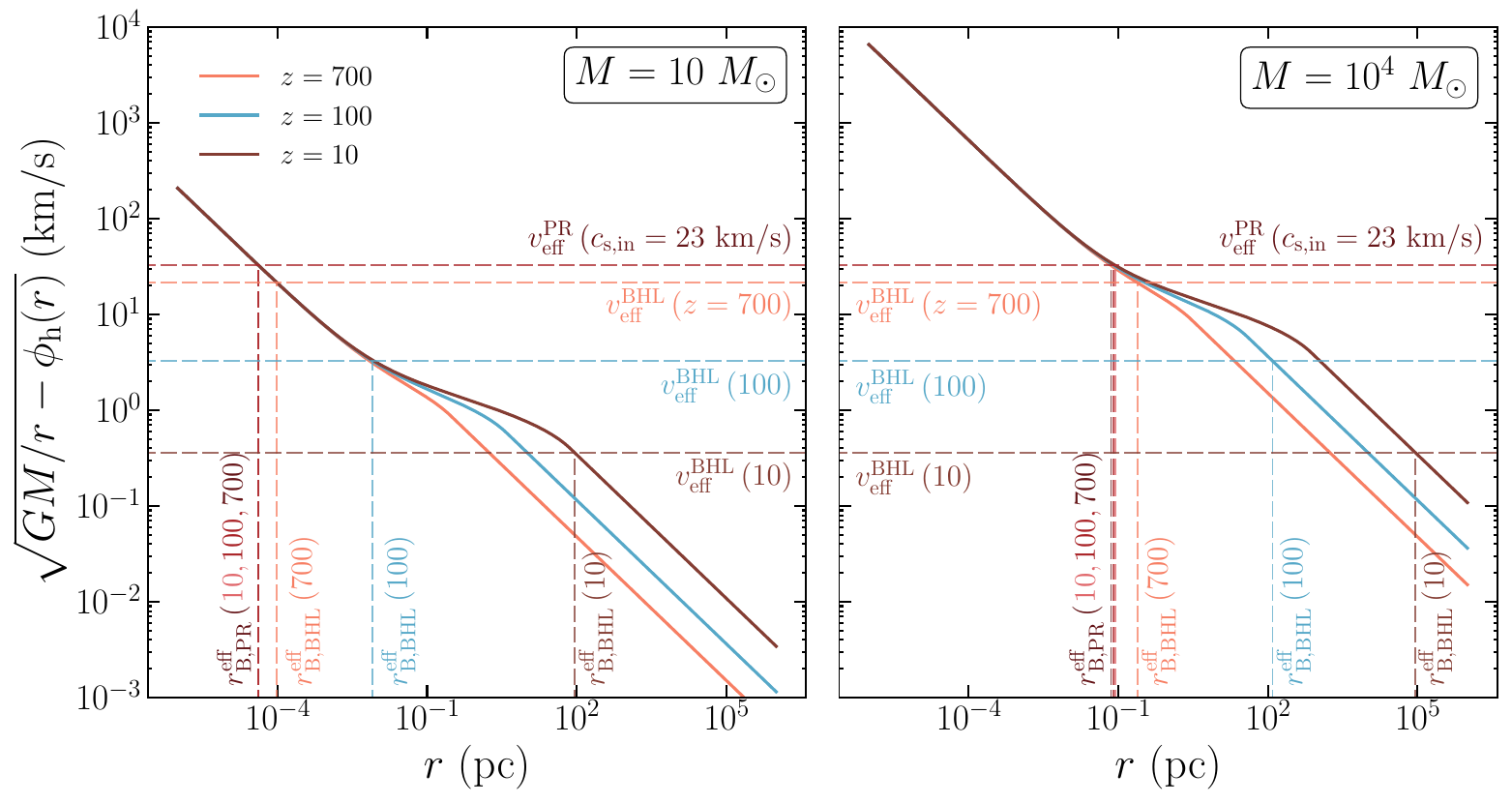} \caption{Graphical solution of Eq.~\ref{eq:effectiveradius}. Solid lines trace the sum of the total ({PBH} plus {DM mini-}halo) potential at different redshifts. Horizontal lines show the values of the effective velocity $\veff$ within the setup of Agius et al.~\protect\cite{us}
The intersection between the horizontal lines and the curves describing the potential determines the value of the effective Bondi radius $\rBeff \left(z \right) $. }
\label{fig:potential}
\end{figure}
As in previous works,~\cite{PR-halos,Serpico} we account for the contribution of DM mini-halos to the gravitational potential by replacing the right-hand side of Eq.~\ref{eq:bondiradius} with the total potential of BH and DM mini-halo. Hence, the generalised Bondi radius $\rBeff $ is obtained solving the following equation
\begin{equation} 
\label{eq:effectiveradius}
\veff ^2 = \frac{G M}{r} - \phi_\mathrm{h}(r)\,,
\end{equation}
where $\phi_\mathrm{h}$ is the gravitational potential of the DM mini-halo (details in Agius et al.~\cite{us}).

The solid lines in Fig.~\ref{fig:potential} trace the total ({PBH} plus {DM mini-}halo) potential, as a function of distance from the PBH. 
Different colours correspond to different redshifts. In the inner region, the total potential is dominated by the PBH contribution and hence follows a $1/r$ slope; in the intermediate region, the contribution of the DM mini-halo becomes increasingly relevant as it builds up over time; finally, beyond the radius of the mini-halo, we recover a $1/r$ potential, whose normalisation now depends on the total mass of {PBH} and DM halo. The horizontal dashed lines correspond to different values of $\veff$. 
The intersection between these and the curves describing the potential determines the value of the effective Bondi radius, $\rBeff $.

Here, the difference between our treatment and previous results lies exclusively in the definition of the effective velocity. Within the PR model, the relevant temperature is that of the ionised gas surrounding the PBH. We set it to $T \sim 10^4$~K, which corresponds to $\veff \sim \cs \sim 20$~km/s ($\veff^\mathrm{PR}$ in the plot). For these values, $\rBeff$ is small and falls in the region where the gravitational potential is dominated by the PBH. As a consequence, the accretion process is almost unaffected by the growth of the DM mini-halo in the outer region.

Previous works,~\cite{Serpico} based on the BHL model, instead computed the effective velocity considering the average properties of the cosmological medium. In this case, as the Universe cools down, the effective velocity ($\veff^\mathrm{BHL}$ in the plot) progressively decreases to below $1$ km/s. Thus, increasingly larger portions of the DM mini-halo are enclosed within $\rBeff$, which eventually encompasses the mini-halo entirely.

\begin{figure}[t]
\centering
\includegraphics[width =.9\columnwidth]{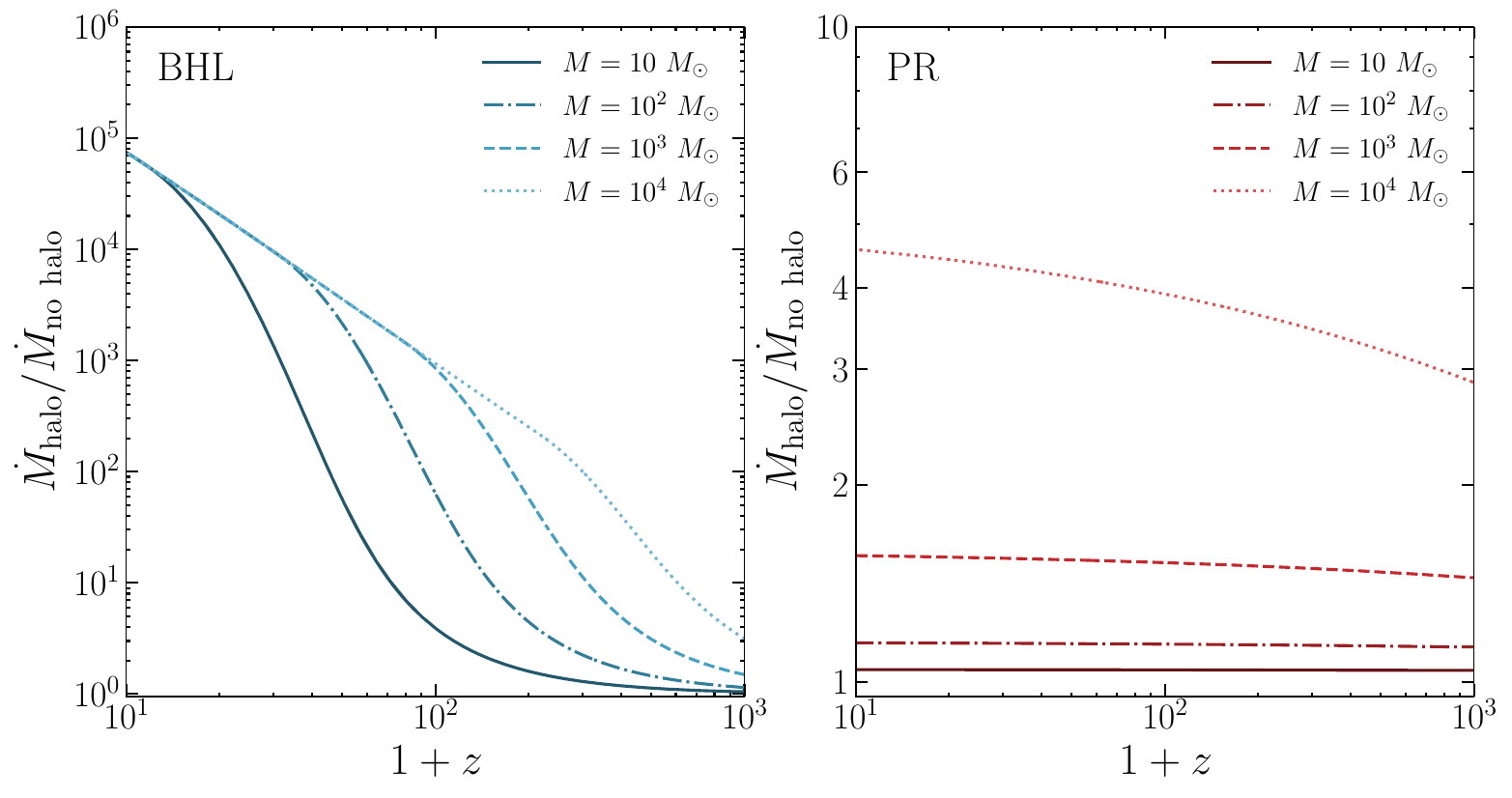}
\caption{Ratio between the accretion rates with and without DM mini-halos. Left: without local heating (BHL). Right: with local heating (PR). Notice the different scaling of the y-axes. Details in Agius et al.~\protect\cite{us} } 
\label{fig:MdotWHalo}
\end{figure}

The DM mini-halo can reach sizes of hundreds of pc, growing up to $\mathcal{O}(10^2)$ times the BH mass at $z\sim 10$. When its contribution is included in the total mass, the accretion rate ($\dot{M}\propto M^2$) is boosted by up to five orders of magnitude.
This can be appreciated in the left panel of Fig.~\ref{fig:MdotWHalo}, displaying the transition from BH-only accretion to a regime dominated by the mini-halo.

Instead, when the local increase of temperature is taken into account, the accretion rate is barely affected by the growth of the DM mini-halo (right panel). 
While our results exploited the PR model to quantify it, such local increase in temperature is expected to occur quite generally as a result of the accretion process (see e.g. Ali-Haïmoud et al.~\cite{Ali-H} and references therein). 

\section{Conclusions}

Local heating of the gas around accreting PBHs can prevent DM mini-halos from playing a relevant role in the accretion process. This suggests that the role of these DM structures in boosting PBH accretion rates may have been significantly overestimated in the literature. 
This issue would benefit from further analysis, possibly including dedicated simulations in the cosmological context.

\section*{Acknowledgments}

I thank the organizers of the Rencontres de Moriond for a fantastic conference. I acknowledge support from the ANR project GaDaMa (ANR-18-CE31-0006).

\end{document}

%% file: macros.tex
\newcommand{\Mdot}{\ensuremath{\dot{M}}}

\newcommand{\fPBH}
{\ensuremath{f_\mathrm{PBH}}}
\newcommand{\veff}{\ensuremath{v_\mathrm{eff}}}

\newcommand{\rB}{\ensuremath{r_\mathrm{B}}}

\newcommand{\rBeff}{\ensuremath{r_\mathrm{B}^\mathrm{eff}}}

\newcommand{\cs}{\ensuremath{c_\mathrm{s}}}

\newcommand{\vrel}{\ensuremath{v_\mathrm{rel}}}

\def\beq{\begin{equation}}
\def\eeq{\end{equation}}
\def\bea{\begin{eqnarray}}
\def\eea{\end{eqnarray}}

%% file: moriond.bbl
\section*{References}

\begin{thebibliography}{99}
\bibitem{us} D. Agius, R. Essig, D. Gaggero, F. S., G. Suczewski and M. Valli, arXiv:2403.18895.

\bibitem{PR} K. Park and M. Ricotti, \Journal{\APJ}{767}{163}{2013}.

\bibitem{Serpico} P. D. Serpico, V. Poulin, D. Inman and K. Kohri, \Journal{\PRR}{2}{023204}{2020}.

\bibitem{Bondi} H. Bondi, \Journal{\em{MNRAS}}{112}{195}{1952}.


\bibitem{HL} F. Hoyle and R. A. Lyttleton, \Journal{\em{Mathematical Proceedings of the Cambridge Philosophical Society}}{35}{405}{1939}.


\bibitem{PR-halos} K. Park, M. Ricotti, P.  Natarajan, T. Bogdanović and J. H. Wise, \Journal{\em{ApJ}}{818}{184}{2016}.


\bibitem{Ali-H} Y. Ali-Haïmoud and M. Kamionkowski, \Journal{\PRD}{95}{043534}{2017}.

\end{thebibliography}
